\documentclass[aps,pra,twocolumn,notitlepage,superscriptaddress,showpacs,longbibliography,10pt]{revtex4-1}

\usepackage{amsmath}    
\usepackage{amsfonts}
\usepackage{bm}
\usepackage{amssymb}
\usepackage{appendix}
\usepackage{graphicx}   
\usepackage{color}      
\usepackage{subfigure}  
\usepackage{comment}
\usepackage{siunitx}

\usepackage{soul}
\usepackage[normalem]{ulem}

\newcommand{\re}{\textrm{Re}}
\newcommand{\im}{\textrm{Im}}
\newcommand{\dgap}{d_{\textrm{break}}}
\mathchardef\mhyphen="2D

\begin{document}

\title{Experimental realization of a Weyl exceptional ring}
\author{Alexander Cerjan}
\email[]{awc19@psu.edu}
\affiliation{Department of Physics, The Pennsylvania State University, University Park, Pennsylvania 16802, USA}
\author{Sheng Huang}
\affiliation{Department of Electrical and Computer Engineering, University of Pittsburgh, Pittsburgh, Pennsylvania 15261, USA}
\author{Kevin P. Chen}
\affiliation{Department of Electrical and Computer Engineering, University of Pittsburgh, Pittsburgh, Pennsylvania 15261, USA}
\author{Yidong Chong}
\affiliation{School of Physical and Mathematical Sciences, Nanyang Technological University, Singapore 637371, Singapore}
\affiliation{Centre for Disruptive Photonic Technologies, Nanyang Technological University, Singapore 637371, Singapore}
\author{Mikael C. Rechtsman}
\affiliation{Department of Physics, The Pennsylvania State University, University Park, Pennsylvania 16802, USA}

\date{\today}

\begin{abstract}
  Weyl points are isolated degeneracies in reciprocal space that are monopoles of the
  Berry curvature. This topological charge makes them inherently robust to Hermitian perturbations
  of the system. However, non-Hermitian effects, usually inaccessible in condensed matter systems,
  are an important feature of photonics systems, and when added to an otherwise Hermitian Weyl material
  have been predicted to spread the Berry charge of the Weyl point out onto a ring of exceptional points,
  creating a Weyl exceptional ring and fundamentally altering its properties.
  Here, we observe the implications of the Weyl exceptional ring
  using real-space measurements of an evanescently-coupled bipartite optical waveguide array by probing its effects on
  the Fermi arc surface states, the bulk diffraction properties, and the output power
  ratio of the two constituent sublattices.
  This is the first realization of an object with topological Berry charge
  in a non-Hermitian system.
\end{abstract}

\maketitle

In recent years, topological phenomena have been extensively explored in both condensed matter physics
and photonics, as these systems can possess exotic states
which realize back-scattering immune transport even in the presence of
disorder \cite{kane_quantum_2005,konig_quantum_2007,haldane_possible_2008,hsieh_topological_2008,raghu_analogs_2008,wang_observation_2009,koch_time-reversal-symmetry_2010,umucalilar_artificial_2011,hafezi_robust_2011,fang_realizing_2012,kraus_topological_2012,kitagawa_observation_2012,rechtsman_photonic_2013,khanikaev_photonic_2013,hafezi_imaging_2013}.
In three dimensions, the simplest class of topologically non-trivial systems are
Weyl materials \cite{wan_topological_2011,yang_quantum_2011,lu_weyl_2013,xu_discovery_2015,lu_experimental_2015,lv_observation_2015,yang_weyl_2015,soluyanov_type-ii_2015,xiao_synthetic_2015,lu_symmetry-protected_2016,chen_photonic_2016,lin_photonic_2016,xiao_hyperbolic_2016,gao_photonic_2016,fang_topological_2016,noh_experimental_2017},
which possess a set of isolated degeneracies in their band structure that are
sources or sinks of Berry flux \cite{berry_quantal_1984}, and are connected by Fermi arc
surface states.
As these Weyl points possess a topological charge, they must be created or annihilated in
sets of at least two, such that the total topological charge in the Brillouin Zone remains zero.
Thus, any isolated Weyl points in a system are protected against
Hermitian perturbations that preserve translational symmetry, which can only change their location in the Brillouin zone.
However, unlike electronic systems, an important feature of photonic systems is their ability to break
Hermiticity through material gain or absorption, as well as radiative outcoupling.
This enables photonic systems to realize phenomena exclusive to non-Hermitian systems,
such as exceptional points, a class of degeneracies where two or more eigenvalues and their associated
eigenvectors coalesce, and the system possesses a non-trivial Jordan block form \cite{kato_perturbation_1995,heiss_exceptional_2004}.
Exceptional points are commonly found in parity-time symmetric systems \cite{bender_pt-symmetric_1999,bender_complex_2002,makris_beam_2008},
and are associated with a wide range of unusual behaviors in topologically trivial optical systems, such
as unconventional reflection and transmission \cite{lin_unidirectional_2011,feng_experimental_2013,peng_parity-time-symmetric_2014},
promoting single mode operation in lasers \cite{hodaei_parity-time_symmetric_2014,feng_single-mode_2014,peng_loss-induced_2014},
novel methods of controlling polarization \cite{lawrence_manifestation_2014,cerjan_achieving_2017,zhou_observation_2018},
and enhancing the Purcell factor of resonant cavities \cite{lin_enhanced_2016,pick_general_2017,pick_enhanced_2017}.

Despite these successes, only in the last few years have the consequences of non-Hermiticity been explored in topologically
non-trivial systems. There is presently an ongoing theoretical effort to fully classify and characterize non-Hermitian topological systems,
which have been found to exhibit a wide range of unexpected behaviors
including anomalous topological winding numbers and breakdowns of
bulk-edge correspondence \cite{lee_anomalous_2016,leykam_edge_2017,hu_exceptional_2017,weimann_topologically_2017,shen_topological_2018,kunst_biorthogonal_2018,yao_edge_2018,gong_topological_2018,li_topological_2018}.
In this context, non-Hermitian Weyl media can serve an exemplary role for understanding
the intersection between non-Hermitian and topological physics, since
they exhibit a set of distinctive behaviors that have been
theoretically predicted \cite{xu_weyl_2017,cerjan_arbitrary_2018,zyuzin_flat_2018}, but have not been previously demonstrated
experimentally. Adding a non-Hermitian perturbation to a Weyl medium is predicted to change the Weyl point into a ring of exceptional
points---a Weyl exceptional ring (WER)---that preserves the
topological charge of the original Weyl point. Although Weyl points
act as magnetic monopoles of Berry curvature, WERs are the first known
non-point-like source of Berry charge.

\begin{figure*}[t!]
  \centering
  \includegraphics[width=0.98\linewidth]{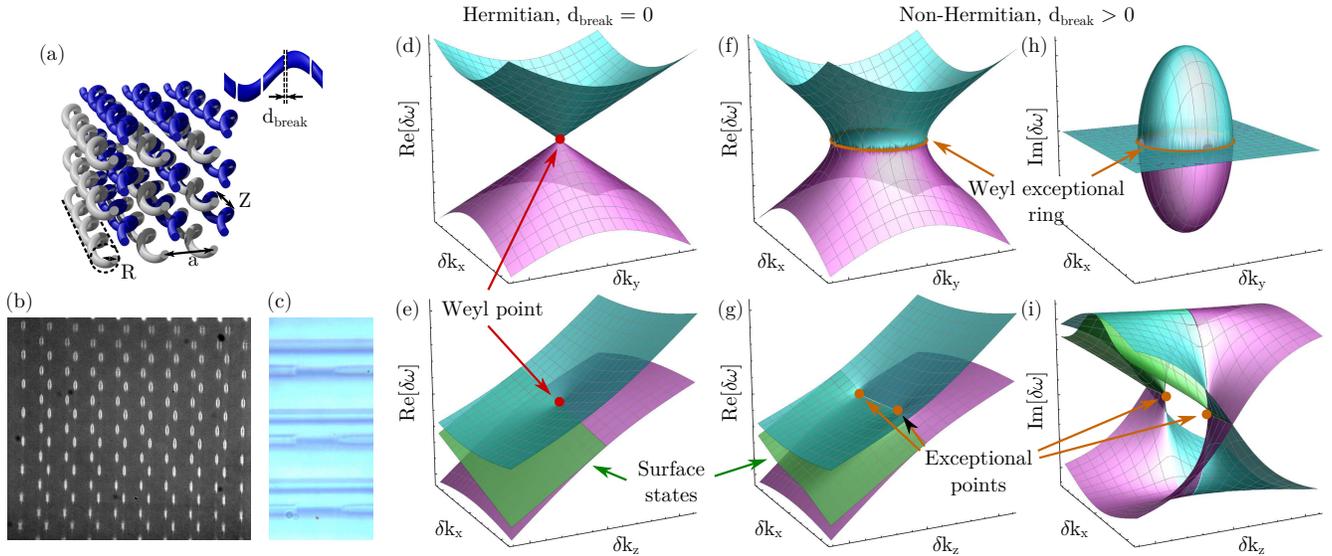}
  \caption{\textbf{Helical waveguide array and corresponding band structure supporting a Weyl exceptional ring.}
    (a) Schematic of the bipartite helical waveguide array in which the rotations of the two sublattices are
    out of phase by a half-cycle and breaks have been added to one of the sublattices. (b) Grayscale microscope image of the output facet of one of the
    helical waveguide arrays.
    (c) Microscope image showing breaks added to the top layer of a helical waveguide array. Within the breaks,
    out of focus waveguides deeper in the array can be seen.
    (d)-(e) Band structures in the $\delta k_x \delta k_y$ and $\delta k_x \delta k_z$ planes
    with $\delta k_y = 0$ and $\delta k_z = 0$, respectively, for a Hermitian waveguide array, $\tau = 0$, showing a Type-II Weyl point. (f)-(g) Similar to (d)-(e), except
    with breaks added to the waveguides, $\tau = 0.2$, so that the band structure possesses a Weyl exceptional ring in the $\delta k_x \delta k_y$ plane
    which is intersected twice by the $\delta k_x \delta k_z$ plane, exhibiting two exceptional points. (h)-(i) Imaginary portion of the band structure
    for the same systems considered in (d)-(e).
    Surface states are shown schematically in (e), (g), and (i) for the states localized to the surface with unbroken waveguides.}
  \label{fig:1}
\end{figure*}

Here we experimentally observe a WER in a 3D photonic lattice consisting of
evanescently coupled single-mode helical waveguides, fabricated using femtosecond direct
laser writing \cite{szameit_discrete_2010}.
To remove the Hermiticity of this system,
we insert breaks into half of the helical waveguides, by periodically
skipping the writing of a specified length of these waveguides, as shown in Figs.\ \ref{fig:1}a-c (see Methods).
Within these breaks the confining potential for the light is removed, resulting
in strong coupling to radiating modes and yielding a tunable mechanism for adding
loss by increasing the length of these breaks.
Thus, by starting with the paraxial wave equation for weakly confined
waveguide modes, and arranging the waveguides in a specific bipartite lattice,
we show that the 3D band structure of this system realizes
the $2 \times 2$ non-Hermitian Weyl exceptional ring Hamiltonian,
\begin{equation}
  \hat{H} \approx v_\perp (\delta k_x \hat{\sigma}_x + \delta k_y \hat{\sigma}_y) + v_z \delta k_z (\hat{I} - |b| \hat{\sigma}_z) + i v_z \tau (\hat{\sigma}_z - |b| \hat{I}), \label{eq:H}
\end{equation}
whose eigenvalues, $\delta \omega$, are the band frequencies relative to the frequency of the underlying Weyl point
for the wavevector components transverse, $\delta \mathbf{k}_\perp = (\delta k_x, \delta k_y)$,
and parallel, $\delta k_z$, to the waveguide axis. The details of this derivation are
included in the Supplementary Information. Here, $\tau$ characterizes the strength of the loss
added to one of the two sublattices of waveguides, $v_\perp$ and $v_z$ are the group velocities
in the transverse and parallel directions, respectively, $\hat{\sigma}_{x,y,z}$ are the Pauli
matrices, $\hat{I}$ is the identity, and $b$ is a dimensionless parameter, with $|b| \ll 1$.

In the Hermitian limit, $\tau = 0$, this helical waveguide array possesses a type-II
Weyl point, whose dispersion is strongly anisotropic because both bands represent modes traveling
in the same direction along the $z$ axis \cite{noh_experimental_2017}.
Here we consider the helical waveguide array as a 3D photonic crystal, not
a 2D system in the paraxial limit, where a Weyl point exists in the $\delta \omega(k_x,k_y,k_z)$ band structure.
Although there is a complementary pair of bands
representing modes traveling in the opposite direction at the same frequency, the weak back-scattering in this system
implies negligible coupling between the forward and backward propagating modes, allowing either pair of bands
to be considered independent of the other.
The distinctive conical band structure of this system at the Weyl point, $\delta \omega = 0$,
is shown in Figs.\ \ref{fig:1}d-e, yielding a large transverse group velocity
at this frequency across nearly the entire transverse Brillouin zone.

\begin{figure*}[t!]
  \centering
  \includegraphics[width=0.80\linewidth]{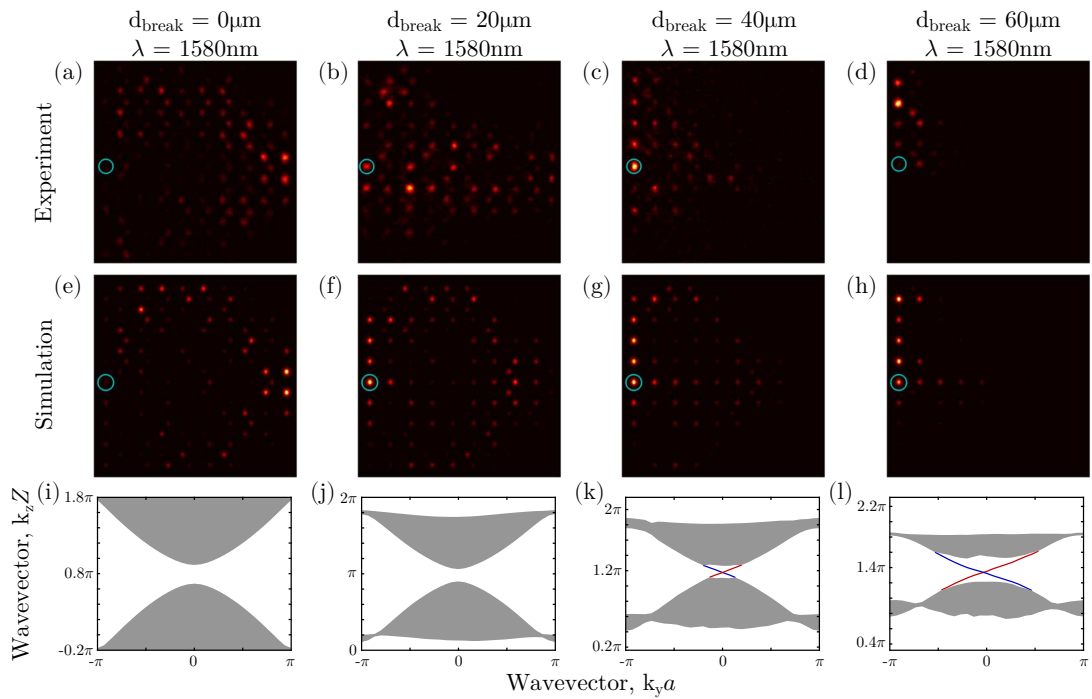}
  \caption{\textbf{Direct observation of a topological transition through the emergence of Fermi arc surface states.}
    (a)-(d) Output intensity plots when light is injected in to a single waveguide at the bottom of the lattice,
    indicated by cyan circle, with a total system length of $L = \SI{8}{\centi\meter}$, at $\lambda = \SI{1580}{\nano\meter}$, for four different break lengths, $\dgap = 0, 20, 40, 60\SI{}{\micro\meter}$.
    This drives the system through a topological transition, and a Fermi arc state is seen in (d).
    (e)-(h) Corresponding full-wave simulation results calculated using the beam propagation method, showing
    good agreement with the experimental results. (i)-(l) Isofrequency contours of a semi-infinite
    helical waveguide array calculated using full wave simulations and a diagonalization procedure \cite{leykam_anomalous_2016}.
    Blue and red curves indicate surface states traveling on the top and bottom of the device, respectively, while
    gray indicates the regions of the bulk bands.}
  \label{fig:surf}
\end{figure*}

However, as loss is added to one sublattice in the bipartite waveguide array by increasing the break
lengths, $|\tau| > 0$,
the two bands begin to merge together starting at the Weyl point, and proceeding radially
outward in the transverse direction, as shown in Fig.\ \ref{fig:1}f. This process yields
a closed contour of exceptional points at $\delta k_z = 0$ between the upper and lower bands
on which the Berry charge of the underlying Weyl point is exactly preserved, resulting in a WER \cite{xu_weyl_2017,cerjan_arbitrary_2018}.
Within this circular region in $\delta \mathbf{k}_\perp$, and for a range of $\delta k_z$ near that of the WER,
the real part of the bands are nearly flat, resulting in an extremely small transverse intensity transport velocity,
which is the non-Hermitian generalization of the group velocity we observe in waveguide arrays \cite{schomerus_non-hermitian-transport_2014}.
These flat bands can be seen by viewing the band structure in the $\delta k_x - \delta k_z$ plane, shown in Fig.\ \ref{fig:1}g
for $\delta k_y = 0$, between the two exceptional points where this plane intersects the WER.
The small transverse intensity transport velocity found in this region is in contrast to the large transverse
group velocity observed at Weyl points, and forms the basis
for one experimental probe of the WER. A second consequence of adding spatially patterned loss,
is that the eigenmodes of the system begin to localize to either the lossless or
lossy waveguides depending on whether they correspond to lossless or attenuating modes,
as the loss produces an effective impedance mismatch between the two sublattices \cite{figotin_dissipative_2012,cerjan_eigenvalue_2016}.
This feature of non-Hermitian systems has been previously observed in parity-time symmetric
optical systems \cite{guo_observation_2009,ruter_observation_2010}, and provides the theoretical basis for a second experimental probe of the WER.
Note that to form a WER, it is critical that the loss is only added to a single sublattice of the system,
which realizes the non-trivial non-Hermitian term $i v_z \tau \hat{\sigma}_z$ in Eq.\ (\ref{eq:H}).
Adding an equal amount of loss to both sublattices would represent a trivial non-Hermitian perturbation
of the form $i v_z \tau \hat{I}$, which preserves the Weyl point, as shown in the Supplementary Information.

One important consequence of the presence of Weyl points in the spectrum of a Hermitian
system is the appearance of Fermi arc surface states at the spatial boundaries of the device.
These surface states form open arcs connecting the projections of pairs of Weyl points with
opposite topological charge in the surface Brillouin zone. When the system becomes
non-Hermitian, the Fermi arc states persist, but now connect the projection of the pair of
WERs which formed from the underlying Weyl points.
Observing the Fermi arc
surface states of the helical waveguide array constitutes the third experimental probe of the WER,
confirming its topological charge.

The specific system that we use to realize a WER consists of a bipartite square lattice
with two waveguides per unit cell, with radius $R = \SI{4}{\micro\meter}$, transverse lattice constant $a = 29\sqrt{2} \SI{}{\micro\meter}$,
and helix period $Z = \SI{1}{\centi\meter}$ in the $z$ direction, as depicted in Fig.\ \ref{fig:1}a. Both waveguides within a unit cell
have clockwise helicity, but their rotational phases are offset by a half-cycle, such
that their nearest neighbor distances change as a function of $z$ \cite{leykam_anomalous_2016}.
A microscope image of a cross-sectional cut of the waveguide array at the output facet
is shown in Fig.\ \ref{fig:1}b.
Finally, 16 evenly-distributed breaks with length $\dgap$ are added to only one of
the two waveguides per unit cell, dramatically increasing its coupling to radiating
modes, and resulting in an effective on-site loss in those waveguides, see Fig.\ S1.
A microscope image of an array of isolated waveguides possessing breaks of different lengths
is shown in Fig.\ \ref{fig:1}c.
Although the spatial distribution of loss in this system resembles that of parity-time symmetric systems,
we note that the helical modulation breaks the inversion symmetry of the system, a necessary
condition for finding Weyl points and WERs, such that the system is not parity-time symmetric.

\begin{figure*}[t!]
  \centering
  \includegraphics[width=0.90\linewidth]{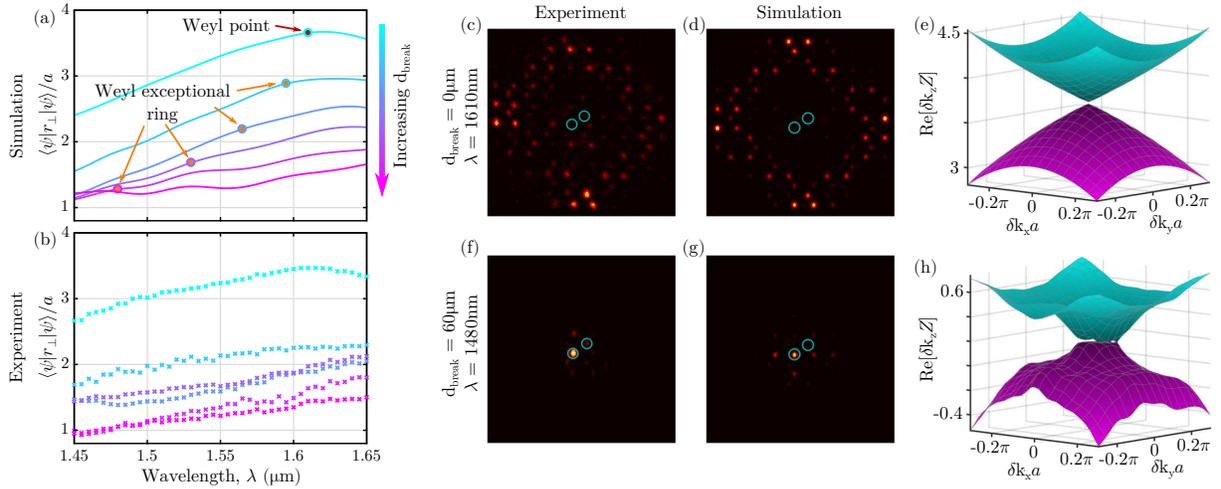}
  \caption{\textbf{Distinguishing a WER from a Weyl point by observing the transverse radial propagation.}
    (a)-(b) Simulations and experimental observations of the transverse radial propagation, $\langle \psi | \mathbf{r}_\perp | \psi \rangle / a$,
    for light injected into the center of the helical waveguide array
    as a function of the injected wavelength for six  different break lengths $\dgap = 0,20,40,50,60,70 \SI{}{\micro\meter}$.
    The Hermitian system is shown in cyan, and redder colors indicate longer break lengths. Wavelengths where simulations
    predict either a Weyl point or WER are indicated in red and orange respectively. (For $\dgap = \SI{70}{\micro\meter}$ the transition occurs near $\lambda = \SI{1400}{\nano\meter}$.)
    (c)-(d) Output intensity plots for light injected into the center of the system at the two indicated waveguides for
    the Hermitian system at the topological transition, $\lambda = \SI{1609}{\nano\meter}$, with a system length of $L = \SI{4}{\centi\meter}$.
    (e) Isofrequency surface for the Hermitian system at the topological transition calculated using full wave simulations and the cut and project method \cite{leykam_anomalous_2016}.
    (f)-(h) Similar to (c)-(e), except for the non-Hermitian system with $\dgap = \SI{60}{\micro\meter}$ at $\lambda = \SI{1480}{\nano\meter}$.
    Note, the roughness seen in the non-Hermitian band structure simulations in (h) is a
    numerical artifact in the diagonalization procedure stemming from the large radiative background when $\dgap > \SI{0}{\micro\meter}$.}
  \label{fig:conical}
\end{figure*}

Given the large disparity between their transverse and longitudinal lattice constants,
helical waveguide arrays are typically analyzed in the paraxial limit to separate these two scales \cite{yariv_optical_1984}.
Then, Maxwell's equations describing the diffraction of light propagating through the array
are reduced to a two-dimensional Schr\"{o}dinger-like equation, in which $z$ acts as a temporal direction,
and the potential confining the light is proportional to the index of refraction of
the waveguides relative to the surrounding index, $\delta n(x,y,z)$.
When Maxwell's equations are
written in this way, the operating frequency becomes an adjustable parameter, while the longitudinal
wavevector component, $k_z$, acts as an effective `energy.' Thus, solutions to the paraxial
equation are isofrequency surfaces of the full three-dimensional band structure. For non-Hermitian
paraxial systems, the amplification or attenuation of a band is instead found as the imaginary
portion of $k_z$, yielding gain or loss per unit length in $z$. As our system contains a Weyl point
or WER at $\delta \omega = 0$, different choices of frequency can result in topologically distinct
two-dimensional band structures of the paraxial equation. In particular, isofrequency surfaces for
$\delta \omega > 0$ are conventional insulators (in the sense that they have a topologically trivial band gap),
while those for $\delta \omega < 0$ are topological insulators.

Using different choices of inputs, we are able to observe three distinct behaviors associated with a WER.
First, we demonstrate that our system exhibits a topological transition by observing the appearance of Fermi arc
states for increasing $\dgap$. Second, we note that
as $\dgap$ is increased, a signal injected into the center of the waveguide array at the topological transition
experiences progressively more localization as the radius of the region of nearly flat bands at the center of the WER
expands. Finally, we observe the ratio of the output power carried on the two sublattices of the
system to demonstrate the spatial localization of the eigenmodes due to the non-uniform distribution
of the loss.

The observation of a topological transition in this helical waveguide array relies on an
additional consequence of creating loss in the system by adding breaks to the waveguides:
light propagating within a break in the lossy waveguide accumulates phase at a slower rate
than light propagating in the lossless waveguide due to the lower index of refraction in
these breaks relative to the index of an unbroken waveguide. As we show in the Supplementary Information,
this difference in phase accumulation shortens the wavelength where the topological transition
due to the Weyl point or WER occurs.
In other words, by fixing the operating wavelength and increasing the break length, the chosen isofrequency
surface can be driven through a topological transition due to the motion of the WER. To observe this topological transition, we inject light
into a single waveguide at the boundary of the lattice and look for the appearance of Fermi arc surface states 
at the output facet of the system. If a surface state is present, light should remain relatively confined
to the system's surface, otherwise it will diffract in to the bulk.
The wavelength is fixed at $\lambda = \SI{1580}{\nano\meter}$, which is less than the wavelength of the Weyl point
in the Hermitian system when $\dgap = \SI{0}{\micro\meter}$, at $\lambda_{\textrm{WP}} = \SI{1609}{\nano\meter}$. Thus, at this
wavelength the injected signal in the Hermitian system simply diffracts, as there is no Fermi arc state
at the operating wavelength. However, as $\dgap$ is increased, the wavelength of the topological
transition at the WER decreases, leading to the appearance of a Fermi arc surface state for $\dgap = \SI{60}{\micro\meter}$.
At this break length, simulations predict that the WER is at $\lambda_{\textrm{WER}} = \SI{1480}{\nano\meter}$.
This process is shown in Fig.\ \ref{fig:surf}.

To demonstrate that the appearance of surface states in the previous experiment is due
to a WER and not a Weyl point, we study the consequences of opening a flat band
region in the center of the Brillouin zone. As the underlying Hermitian system with $\dgap = \SI{0}{\micro\meter}$
possesses a type-II Weyl point, its isofrequency surface at this point is conical, leading
to a large transverse group velocity, as shown in Fig.\ \ref{fig:conical}e. However, for either
shorter or longer wavelengths the isofrequency surfaces are hyperbolic. Thus, when light is injected
into the center of the structure for propagation distances in $z$ short enough so that the beam 
does not reflect off the boundary of the system, the Weyl point is
seen as a peak in the transverse radial expectation value, $\langle \psi | \mathbf{r}_\perp | \psi \rangle / a$.
In the cyan curve of Fig.\ \ref{fig:conical}a, which corresponds to the Hermitian waveguide array,
the peak of the experimentally observed transverse radial expectation value
is in close agreement with the numerically predicted location of the Weyl point at $\lambda = \SI{1609}{\nano\meter}$.
But, as the waveguide break length is increased, shifting the location of the topological transition
to shorter wavelengths, the peak in the transverse radial expectation value disappears,
indicating that there is no significant difference in this quantity between wavelengths where
the topological transition occurs and wavelengths with hyperbolic dispersion.
This demonstrates that the system experiences a topological transition
without a conventional band touching at a Weyl point, and as such is unlike previously
observed topological transitions.
Our observation is consistent with the formation of a WER in the helical waveguide array,
and inconsistent with the existence of an ordinary Weyl point, as a WER
flattens the center of the isofrequency surface in the Brillouin zone and decreases the
transverse intensity transport velocity so that the intensity transport velocity profile is similar to those found
in hyperbolic isofrequency surfaces away from the topological transition.

\begin{figure}[t!]
  \centering
  \includegraphics[width=0.62\linewidth]{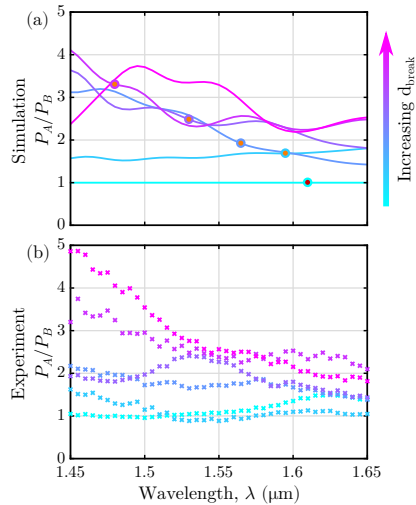}
  \caption{\textbf{Output power ratio as a signature of a WER.}
    (a)-(b) Simulations and experimental observations of the sublattice power ratio, $P_A / P_B$,
    for light injected into the center of the helical waveguide array
    as a function of the injected wavelength for six different break lengths $\dgap = 0,20,40,50,60,\SI{70}{\micro\meter}$.
    The Hermitian system is shown in cyan, and redder colors indicate longer break lengths. Wavelengths where simulations
    predict either a Weyl point or WER are indicated in red and orange respectively.}
  \label{fig:avb}
\end{figure}

A second confirmation that the broken helical waveguide array possesses a
WER can be seen in the distribution of the output power of the device between the
two sublattices of the system. One consequence of adding spatially inhomogeneous loss to a system
is that the system's eigenmodes localize to either the lossless or lossy regions.
(In the limit of very strong loss, this localization can be proven to be perfect \cite{figotin_dissipative_2012,cerjan_eigenvalue_2016}.)
The localization of the eigenmodes is reflected in their respective eigenvalues, which either correspond
to nearly lossless or strongly attenuated propagation. This effect can be viewed as the
result of an impedance mismatch between the different sublattices of the system due to
the spatially inhomogeneous loss, and can lead to loss-induced transmission in waveguides \cite{guo_observation_2009}
and reverse pump dependence in lasers \cite{liertzer_pump-induced_2012,brandstetter_reversing_2014,peng_loss-induced_2014}. In contrast, the eigenmodes of a Hermitian system whose
elements all have the same impedance, i.e.\ index of refraction, are evenly distributed
over the entire system. Thus, for light injected into a Hermitian bipartite waveguide array
whose constituent elements all have the same index of refraction, the output power should be
evenly distributed over the two sublattices of the system, yielding an output power ratio
$P_A / P_B \approx 1$, where $A$ and $B$ denote the two sublattices of the system.
However, as loss is added to the $B$ sublattice, the propagating modes with nearly lossless transmission
are localized to the lossless waveguides, instead resulting in $P_A / P_B \gg 1$.
This analysis is confirmed in both simulation and experiment in Fig.\ \ref{fig:avb}, the Hermitian
system has an output power ratio near $1$, but as the break length is increased this quantity
begins to diverge. This final experiment should be viewed as direct confirmation that we have achieved the non-trivial non-Hermitian term $i \tau \sigma_z$ in Eq.\ (\ref{eq:H}) which results in
the formation of a WER.

In conclusion, we have observed a helical waveguide array supporting a WER at optical frequencies
by adding breaks to half of the waveguides, breaking the Hermiticity of the system.
As we have shown, non-Hermitian perturbations yield a fundamentally new class of topological objects,
WERs, in contrast with Hermitian perturbations
which simply shift a Weyl point's location in the Brillouin zone.
Although exceptional rings have been previously observed in topologically trivial
systems \cite{zhen_spawning_2015}, this
experiment not only provides the first observation of a non-point-like source of Berry charge, but also
directly demonstrates that topological charge and Fermi arc surface states are preserved
in the presence of a non-Hermitian perturbation to the system,
even as the Weyl point itself transforms into a WER.
This experimental confirmation
of analytic predictions of the properties of non-Hermitian Weyl materials paves the way
towards additional theoretical and experimental studies, in particular understanding the complex interplay
between band topology and the non-trivial topological structure associated with exceptional points.

\section*{Methods}

The helical waveguide arrays are written using a titanium:sapphire laser and amplifier system (Coherent:RegA 9000) with pulse duration $\SI{270}{\femto\second}$,
repetition rate $\SI{250}{\kilo\hertz}$, and pulse energy $\SI{880}{\nano\joule}$, into Corning Eagle XG borosilicate glass with a refractive index of $n_0 = 1.473$.
The laser writing beam is sent through a beam-shaping cylindrical telescope to control the size and shape of the focal volume, and is
then focused inside the glass chip using a $\times 50$, aberration-corrected microscope objective ($\textrm{NA} = 0.55$).
A high-precision three-axis Aerotech motion stage (model ABL20020) is used to translate the sample during fabrication.
The waveguide breaks are formed by turning off the laser writing beam using AOM (acousto optical modulator) while the motion stage continues to move, and then turning
the beam back on after the desired distance is reached.
Experiments are performed by butt-coupling a single-mode optical fiber
to waveguides at the input facet of the chip, which subsequently couples to the waveguide array. The input light is
supplied by a tunable mid-infrared diode laser (Agilent 8164B), which can be tuned through the $1450\mhyphen\SI{1650}{\nano\meter}$ wavelength
range. After a total propagation distance of $4$ or $\SI{8}{\centi\meter}$ within the array, depending on the experiment,
the light output from the waveguide array is observed using a $0.2$ NA microscope objective lens and a near-infrared InGaAs camera (ICI systems).

\section*{Data and code availability}
The data and code that support the findings of this study are available from the corresponding authors on reasonable request.

\begin{acknowledgments}
  The authors thank Jiho Noh for discussions about experimental technique.
  M.C.R.\ and A.C.\ acknowledge support from the National Science Foundation under grant numbers
  ECCS-1509546 and DMS-1620422 as well as the Charles E. Kaufman foundation under grant number KA2017-91788.
  K.C.\ and S.H.\ acknowledge the National Science Foundation under
  Grants No. ECCS-1509199 and No. DMS-1620218.
  C.Y.D.\ is supported by the Singapore MOE Academic Research Fund Tier 2
  Grants MOE2015-T2-2-008 and MOE2016-T2-1-128, and the Singapore MOE
  Academic Research Fund Tier 3 Grant MOE2016-T3-1-006.
\end{acknowledgments}


%

\newpage

\onecolumngrid

\section*{Supplemental information for: Experimental realization of a Weyl exceptional ring}
\maketitle

\setcounter{equation}{0}
\renewcommand{\theequation}{S{\arabic{equation}}}
\setcounter{figure}{0}
\renewcommand{\thefigure}{S{\arabic{figure}}}

\section{Derivation of the 3D Weyl exceptional ring Hamiltonian from the waveguide Floquet Hamiltonian \label{sec:1}}

As discussed in the main text, the steady-state diffraction of light at a specific frequency
propagating through a waveguide array is described by the paraxial approximation to Maxwell's equations.
For a given electric field amplitude with a fixed linear polarization, $\mathcal{E}(x,y,z)e^{-i \omega t}$, the slowly-varying
envelope, $\psi(x,y,z)$, is
\begin{equation}
  \mathcal{E}(x,y,z) = \psi(x,y,z)e^{i k_0 z},
\end{equation}
in which $k_0 = n_0 \omega / c$, and $n_0$ is the index of refraction of the background material in the system,
and the propagation is assumed to be predominantly along the $z$ axis.
In the limit that the back scattering in the $z$ direction due to the waveguides is negligible, $\psi(x,y,z)$ satisfies the 2D Schr{\" o}dinger
equation,
\begin{equation}
  i \frac{\partial \psi}{\partial z} \approx \hat{H} \psi(x,y,z) = \left[ -\frac{1}{2k_0} \nabla_\perp^2 - \frac{\omega}{c} \delta n(x,y,z) \right] \psi(x,y,z), \label{eq:s2}
\end{equation}
where $\nabla_\perp^2$ is the Laplacian in the transverse directions, $x,y$, and $\delta n = n - n_0$ is the shift in
the refractive index which defines the waveguides. As the helical waveguides are periodic in $z$, with period $Z$,
such that $\delta n(x,y,z + Z) = \delta n(x,y,z)$, Eq.\ (\ref{eq:s2}) is a Floquet problem, whose eigenstates
satisfy
\begin{align}
  \psi(x,y,z+Z) &= e^{-i \beta Z} \psi(x,y,z) \\
  \hat{H}_F \psi &= \beta \psi \label{eq:s4}
\end{align}
in which $\hat{H}_F$ is the Floquet Hamiltonian, and $\beta$ is the Floquet quasi-energy. The Floquet
Hamiltonian is defined using the $z$-evolution operator over a single period,
\begin{equation}
  e^{-i \hat{H}_F Z} = \mathcal{T} e^{-i \int_0^Z \hat{H}(x,y,z) dz},
\end{equation}
while the Floquet quasi-energy corresponds to a slight shift in the wavevector along the axis of the
waveguides,
\begin{equation}
  \beta = k_z - k_0.
\end{equation}

Usually, when solving for the band structure of a photonic crystal, the allowed frequencies are the eigenvalues
of an equation dependent upon the wavevector components of the system, $k_x,k_y,k_z$. However,
in re-writing Maxwell's equations using the paraxial approximation in Eq.\ (\ref{eq:s2}), we have effectively
exchanged the wavevector component along the propagation axis, $k_z$, for the frequency, $\omega$, so that
the exact momentum along the waveguides' axis is now determined by the three degrees of freedom $k_x,k_y,\omega$.
Thus, when $\hat{H}_F$ is non-Hermitian, the propagation constant will develop an imaginary component,
$\beta \in \mathbb{C}$, corresponding to the amplification, $\im[\beta] < 0$, or attenuation, $\im[\beta] > 0$, of the corresponding eigenstate,
so that the total evolution of the electric field can be written as
\begin{equation}
  E(x,y,z;t) = \psi(x,y,z)e^{i (k_0 + \beta) z - i \omega t},
\end{equation}
with $\omega \in \mathbb{R}$.
This amplification or attenuation can instead be considered as an imaginary component of the frequency
using the mode-dependent group velocity in the propagation direction, $v_{g,z}^{(n)}$, as $\Delta k_z = - \Delta \omega / v_{g,z}^{(n)}$ \cite{joannopoulos},
so that the total evolution
of the electric field can also be written as,
\begin{equation}
  E(x,y,z;t) = \psi(x,y,z)e^{i (k_0 + \tilde{\beta}) z - i \tilde{\omega} t},
\end{equation}
with $\tilde{\beta} = \re[\beta]$, and $\tilde{\omega} = \omega - i v_{g,z}^{(n)} \im[\beta]$.

The bipartite square waveguide lattice we consider here can be tuned to a topological transition at
frequency $\omega_0$, as demonstrated in the main text in Fig.\ 2. Thus, to lowest order in the frequency
detuning, $\delta \omega = \omega - \omega_0$, in the neighborhood of the topological transition,
the Floquet Hamiltonian is described by a non-Hermitian Dirac Hamiltonian,
\begin{equation}
  \hat{H}_F \approx v_d(\delta k_x \hat{\sigma}_x + \delta k_y \hat{\sigma}_y) + b \frac{n_0 \delta \omega}{c} \hat{\sigma}_z - i \tau \hat{\sigma}_1 - \Delta \varepsilon \tilde{\tau} \hat{\sigma}_1, \label{eq:s3}
\end{equation}
in which $\tilde{\tau}$ is a real number that parameterizes the loss added to the system through the breaks in one sublattice of
the system, $\Delta \varepsilon$ is the effective shift in the on-site energy due to the reduction
in the average index of refraction from the breaks in the waveguides per unit loss added by the breaks (hence, $\Delta \varepsilon \tilde{\tau}$ is the total decrease
in the on-site energy for a given break length), $\hat{\sigma}_1 = (1/2)(\hat{\sigma}_z + I)$ is the Pauli
matrix for one sublattice of the system, and $v_d,b$ are real constants.
When $\tilde{\tau} = 0$, this Floquet Hamiltonian becomes the traditional Hermitian Dirac Hamiltonian with
an effective mass determined by the frequency, with the topological transition occurring at $\delta \omega = 0$.
But, when $\tilde{\tau} > 0$, the system develops both an on-site loss and on-site energy shift in one
of the two sublattices.

To prove that this Floquet Hamiltonian is equivalent to the non-Hermitian Weyl Hamiltonian in Eq.\ (1) of
the main text, we must re-arrange Eq.\ (\ref{eq:s3}) into a form where $\delta k_z$ appears as a parameter and that
generates the eigenvalues $\delta \omega$. Noting that
\begin{equation}
  \beta = \delta k_z - \frac{n_0}{c} \delta \omega, 
\end{equation}
where $\delta k_z = k_z - n_0 \omega_0 / c$, we can rewrite Eq.\ (\ref{eq:s4}) using Eq.\ (\ref{eq:s3}) as
\begin{equation}
  \left[v_d(\delta k_x \hat{\sigma}_x + \delta k_y \hat{\sigma}_y) - \delta k_z \hat{I} - \frac{\tilde{\tau}}{2}(i + \Delta \varepsilon) (\hat{\sigma}_z + \hat{I}) \right] \psi
  = -(\hat{I} + b \hat{\sigma}_z)\frac{n_0 \delta \omega}{c} \psi.
  \label{eq:s5}
\end{equation}
Equation (\ref{eq:s5}) has the form of a generalized eigenvalue problem. To convert it to an ordinary eigenvalue problem, we seek to factorize
the operator on the right-hand side of the equation and rescale the eigenstate vectors as,
\begin{align}
  \varphi =& \hat{\mathcal{W}} \psi, \\
  \hat{\mathcal{W}}^2 \equiv& -\frac{n_0}{c}(\hat{I} + b \hat{\sigma}_z),
\end{align}
which leads to
\begin{align}
  \hat{H}' \varphi &= \delta \omega \varphi, \\
  \hat{H}' &= \hat{\mathcal{W}}^{-1} \left[v_d(\delta k_x \hat{\sigma}_x + \delta k_y \hat{\sigma}_y) - \delta k_z \hat{I} - \frac{\tilde{\tau}}{2}(i + \Delta \varepsilon) (\hat{\sigma}_z + \hat{I}) \right] \hat{\mathcal{W}}^{-1}. \label{eq:s6}
\end{align}
Assuming that $|b| < 1$, one can directly verify that the appropriate re-scaling operators are
\begin{align}
  \hat{\mathcal{W}} &= i \left(\frac{n_0}{2 c}(1 - \sqrt{1 - b^2}) \right)^{1/2} \hat{I} + i \left(\frac{n_0}{2 c}(1 + \sqrt{1 - b^2}) \right)^{1/2} \hat{\sigma}_z, \\
  \hat{\mathcal{W}}^{-1} &= \frac{c}{n_0\sqrt{1 - b^2}}\left[ i \left(\frac{n_0}{2 c}(1 - \sqrt{1 - b^2}) \right)^{1/2} \hat{I} - i \left(\frac{n_0}{2 c}(1 + \sqrt{1 - b^2}) \right)^{1/2} \hat{\sigma}_z \right],
\end{align}
the same as was previously reported for the Hermitian version of this system \cite{noh_experimental_2017}. Equation (\ref{eq:s6}) can
now be rewritten as
\begin{equation}
  \hat{H}' = \frac{c v_d}{n_0 (1-b^2)}(\delta k_x \hat{\sigma}_x + \delta k_y \hat{\sigma}_y) + \frac{c}{n_0(1-b^2)} \delta k_z (\hat{I} - |b| \hat{\sigma}_z) +\frac{\tilde{\tau}(i + \Delta \varepsilon)}{2}
  \left(\frac{c(1-|b|)}{n_0 (1-b^2)} \right) \left[\hat{\sigma}_z + \hat{I} \right],
\end{equation}
which can be simplified by defining
\begin{align}
  v_\perp &= \frac{c v_d}{n_0 (1-b^2)} \\
  v_z &= \frac{c}{n_0(1-b^2)} \\
  \tau &= \frac{\tilde{\tau}(1-|b|)}{2}
\end{align}
to
\begin{equation}
  \hat{H}' = v_\perp(\delta k_x \hat{\sigma}_x + \delta k_y \hat{\sigma}_y) + v_z \delta k_z (\hat{I} - |b| \hat{\sigma}_z) + v_z \tau (i + \Delta \varepsilon)
  (\hat{\sigma}_z + \hat{I}). \label{eq:sW}
\end{equation}
This is the non-Hermitian Weyl Hamiltonian. Written in this manner, it is clear that $\Delta \varepsilon \tau \hat{\sigma}_z$ is
a Hermitian perturbation to the system which has the effect of changing the location of the topological transition in $k_z$,
while terms proportional to $\hat{I}$ amount to an overall shift in the spectrum of
the system and do not play a role in the formation of either the Weyl point (when $\tau = 0$) or the Weyl exceptional ring (when $\tau > 0$).
As such, these terms have been omitted from Eq.\ (1) of the main text, although the effect of $\Delta \varepsilon \tau \hat{\sigma}_z$
is discussed in the context of Fig.\ 2 of the main text, and in more detail in Sec.\ \ref{sec:4} of the supplementary information.

Finally, to justify that $|b| < 1$ for this system, we provide an analysis of its order of magnitude. Recall from
Eq.\ (\ref{eq:s3}) that $b$ scales the effective mass in the Dirac Hamiltonian. The unbroken Hermitian system
exhibits a Weyl point at $\lambda_W = 1609$nm. If the system is detuned to $\lambda_2 = 1564$nm, the isofrequency
surfaces split at $k_x = k_y = 0$, with $\Delta k_z Z \approx \pi/4$, with $\Delta k_z$ being the splitting between the upper and lower bands. This split is due to the effective mass
in the system, i.e.\ $b n_0 \delta \omega / c = \pi/8Z$. Using the identities $k_0 = n_0 \omega / c = 2 \pi / \lambda$,
and $\delta \omega = (2 \pi c / n_0)(\lambda_2^{-1} - \lambda_W^{-1})$, one can estimate $b = 3.5 \cdot 10^{-5}$.

Throughout our analysis, we have focused only on those modes which are localized within, and propagate along, the single mode
waveguides which comprise the helical waveguide array. When breaks are added to the waveguides, these waveguide modes
couple to `unbound' modes of the system that exist predominantly in the space between the waveguides.
Since we only monitor the light localized to the bound modes of the waveguides at the output facet,
we observe an effective loss rate for the broken waveguides. When solving for the three-dimensional band structure $\omega(k_x,k_y,k_z)$,
this effective loss results in presence of the WER, despite the fact that the perturbation associated with the breaks between waveguides is purely dielectric and requires no absorptive medium.

\section{Simulation parameters}

The simulations presented in this work are performed using the beam propagation method (also called the split-step method) \cite{taha_analytical_1984,agrawal_nonlinear_2012},
which directly evolves the electric field envelope $\psi(x,y,z)$ in the propagation direction ($z$)
in accordance with Eq.\ (\ref{eq:s2}). Band structures shown in Figs.\ 2 and 3 are propagated for a
single helix pitch length $Z$, and then calculated using a diagonalization procedure \cite{leykam_anomalous_2016} (see the supplemental material of that work).
The index variation of the waveguides is modeled as a hyper-Gaussian,
\begin{equation}
  \delta n(x,y,z) = \Delta n e^{[(x^2/\sigma_x^2) + (y^2/\sigma_y^2)]^3},
\end{equation}
for which the lengths of the axes of the waveguide's elliptical cross-section are $\sigma_x = \SI{3.2}{\micro\meter}$ and
$\sigma_y = \SI{4.9}{\micro\meter}$, the background index of refraction
is $n_0 = 1.473$, and the index shift of the waveguides is $\Delta n = 2.6\cdot 10^{-3}$.
For ease of reproduction we repeat the remainder of the system parameters here. The transverse lattice constant is
$a = 29\sqrt{2}\SI{}{\micro\meter}$. The pitch of the helical waveguides is $Z = \SI{1}{\centi\meter}$. The
rotation radius of the helical waveguides is $R = \SI{4}{\micro\meter}$. 

As breaks are added to the system to induce the formation of the Weyl exceptional ring, there are 16
breaks added per helix pitch, equally spaced, each with length $\dgap$, where $\delta n(x,y,z) = 0$.
Thus, for a system with
$\dgap = \SI{60}{\micro\meter}$, a total length of $\SI{960}{\micro\meter}$ has been removed
per $Z$. These breaks are arranged such that the locations in $z$ of closest approach between neighboring waveguides
are always unbroken, and bisect the distance between adjacent breaks.

\section{Loss as a function of break length}

\begin{figure*}[h]
  \centering
  \includegraphics[width=0.30\linewidth]{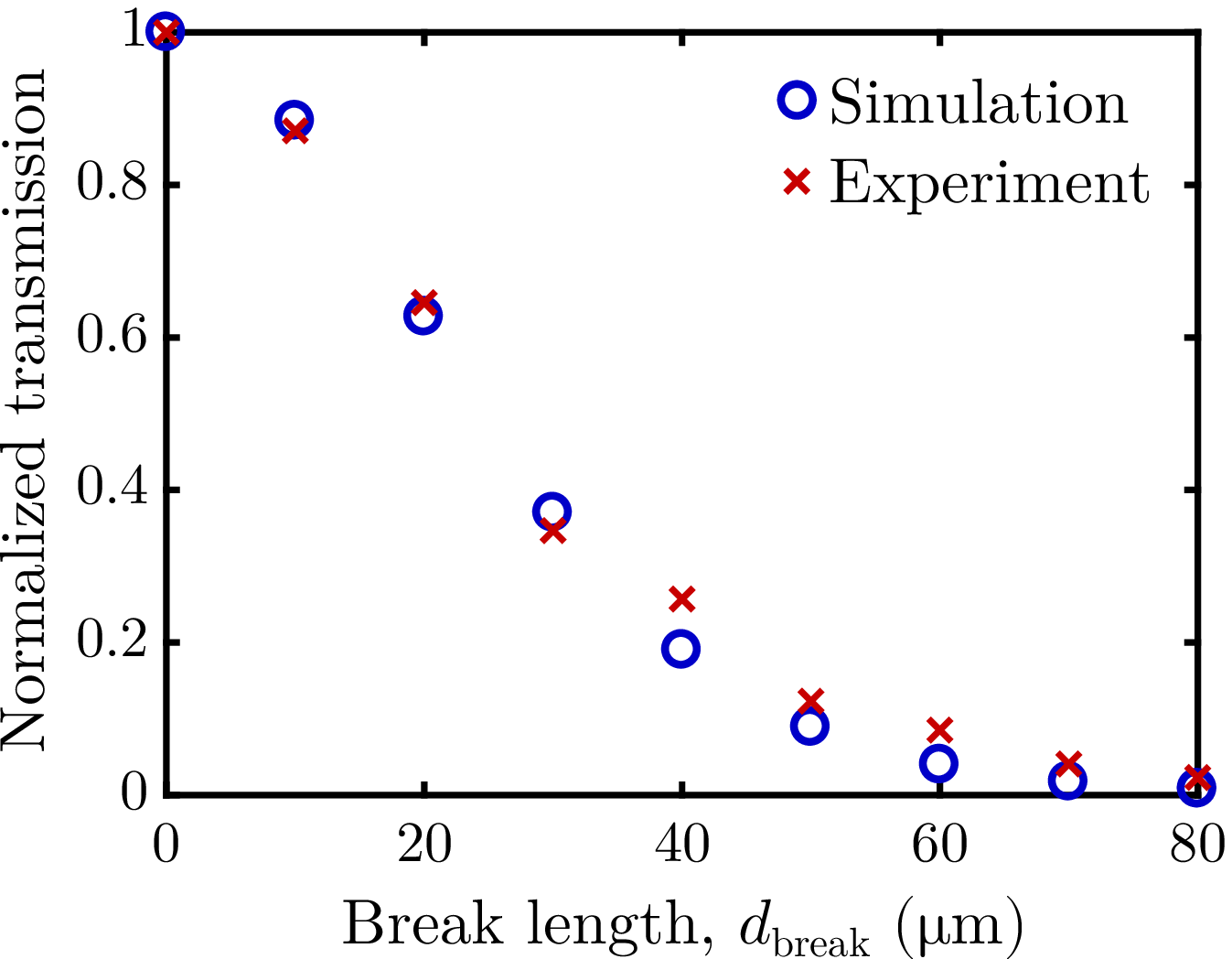}
  \caption{\textbf{Transmission as a function of break length for isolated, straight waveguides.} Experimental results
    are shown as red crosses, and simulation results are shown a blue circles.
    The total length of the system is $L = \SI{4.9}{\centi\meter}$. For the purposes of break placement, the straight
    waveguides are assumed to have a fictitious helix pitch of $Z = \SI{1}{\centi\meter}$, and 16 breaks are placed
    per helix pitch, each with length $\dgap$.}
  \label{fig:loss}
\end{figure*}

To demonstrate that these simulation parameters yield results which agree with the experiment,
we compare the total transmission as a function of break length for isolated, straight waveguides.
As can be seen in Fig.\ \ref{fig:loss}, the chosen simulation parameters faithfully reproduce
the experimental results.

\section{Effect of changing the index of one sublattice \label{sec:4}}

As discussed in the main text, and above in Sec.\ \ref{sec:1}, one of the consequences of adding breaks to one
sublattice of the helical waveguides which comprise the bipartite system is that light propagating through a
break accumulates phase at a slower rate than light propagating through the corresponding region in the unbroken
waveguide. Moreover, as shown in Eq.\ (\ref{eq:sW}), this can result in a change in the frequency, $\delta \omega$,
where the topological transition occurs through the term $\Delta \varepsilon \tau$.
For our system, this effective shift in the index of refraction of one of the sublattice waveguides
is a non-negligible effect. For example, noting
that there are 16 equally long breaks per helix pitch, if $\dgap = \SI{60}{\micro\meter}$, the total length
of the broken region per helix pitch is $\SI{960}{\micro\meter}$, which is $9.6\%$ of the total helix ($Z = \SI{1}{\centi\meter}$).
Thus, it is important to distinguish which features of this system are a consequence of this effective detuning
of the indices of refraction of the two sublattices, a completely Hermitian phenomenon, and which features
are a consequence of the loss added to the system through the inclusion of these waveguide breaks.

\begin{figure*}[ht]
  \centering
  \includegraphics[width=0.70\linewidth]{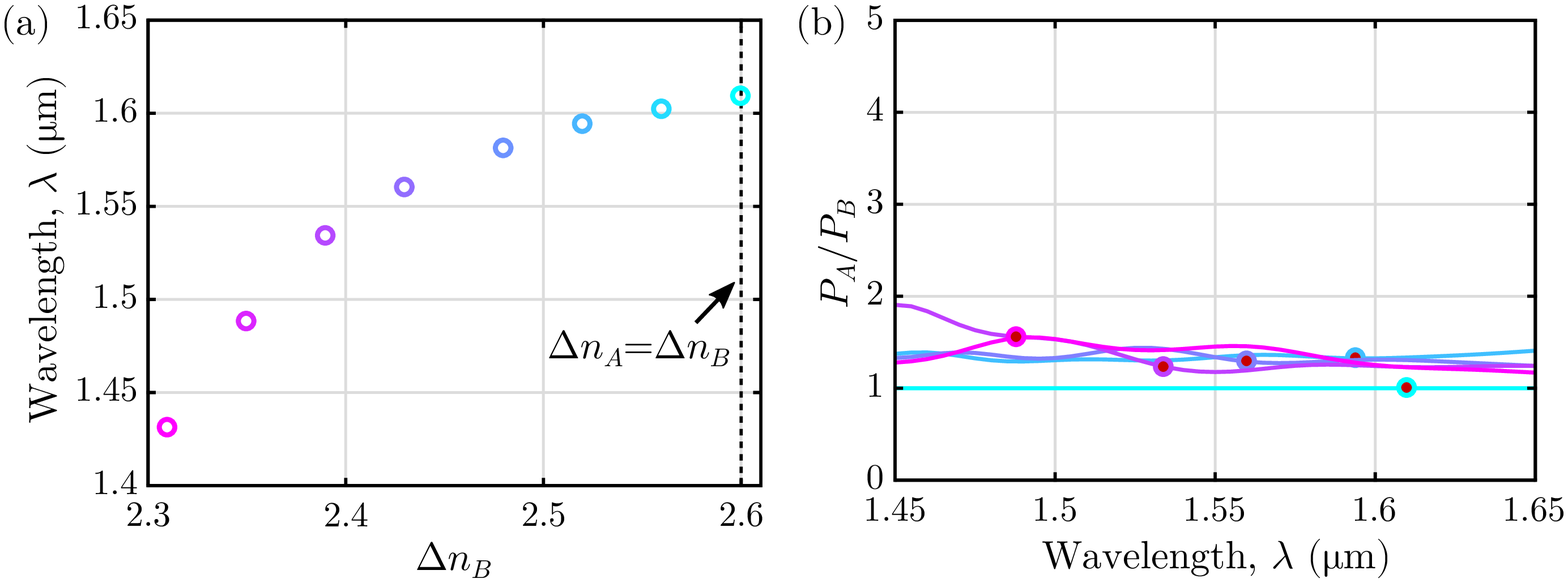}
  \caption{\textbf{Properties of detuned, Hermitian helical waveguides.} (a) Wavelength of the Weyl point
    as the index detuning between the two sublattices is decreased found using full wave simulations.
    The values of $\Delta n_B$ chosen correspond
    to the effective indices of refraction for $\dgap = 0, 10, 20, 30, 40, 50, 60, \SI{70}{\micro\meter}$.
    (b) Full wave simulations of the output power ratio between the two sublattices of the detuned Hermitian system
    with $\Delta n_B = 2.6 \cdot 10^{-3}, 2.52 \cdot 10^{-3}, 2.43 \cdot 10^{-3}, 2.39 \cdot 10^{-3}, 2.35 \cdot 10^{-3}$ (cyan to magenta),
    and $\Delta n_A = 2.6 \cdot 10^{-3}$. The chosen detunings correspond to $\dgap = 0, 20, 40, 50, 60, \SI{70}{\micro\meter}$.}
  \label{fig:dn}
\end{figure*}

To disentangle the effects of the added loss from those of the index detuning, here we study systems which
have the two sublattice indices detuned, but no added breaks. For example, a lossy waveguide with
$\dgap = \SI{60}{\micro\meter}$ and $\Delta n_B = 2.6\cdot 10^{-3}$ has an effective index
of $\Delta n_B = 2.35\cdot 10^{-3}$ if the waveguide were unbroken.
In Fig.\ \ref{fig:dn}(a), we show that the wavelength of the Weyl point decreases in detuned
unbroken Hermitian waveguide arrays with $\Delta n_B < \Delta n_A = 2.6\cdot 10^{-3}$.
As such, we reiterate here that the change in the wavelength of
the topological transition observed in Fig.\ 2 of the main text is a consequence of the detuning of the indices
of refraction of the two waveguides due to the added breaks in one sublattice, and not due to the loss added to
the system through the breaks.

However, the purely Hermitian change in the location of the topological transition
is inconsistent with the later results in the main text. In particular, the output power ratio, $P_A / P_B$
remains near unity for these detuned Hermitian systems, as shown in Fig.\ \ref{fig:dn}(b), which is
in sharp contrast to what is observed for the non-Hermitian systems in Fig.\ 4.
As such, one can conclude that while the wavelength of the topological transition is mostly determined
by the effective index detuning caused by the added breaks, the added loss to one sublattice of the system
has caused this detuning to be a Weyl exceptional ring, and cannot be just a Weyl point.

\section{Adding homogeneous loss: Breaks added to both sublattices}

\begin{figure*}[ht]
  \centering
  \includegraphics[width=0.65\linewidth]{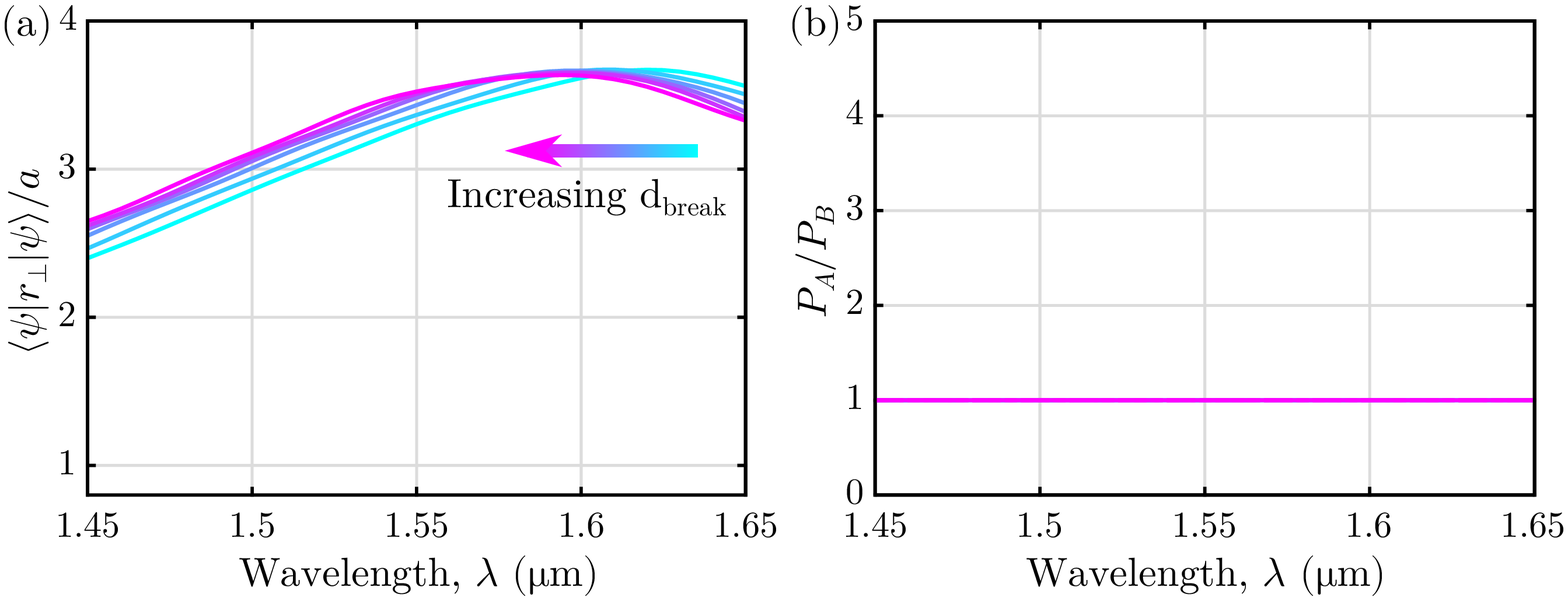}
  \caption{\textbf{Non-Hermitian waveguide arrays with breaks added to both sublattices.}
    (a) Full wave simulations of transverse radial propagation and (b) output power ratio as a function
    of the wavelength for $\dgap = 0, 10, 20, 25, 30, \SI{35}{\micro\meter}$.}
  \label{fig:ddrop}
\end{figure*}

In the previous section, we demonstrated that simply considering the Hermitian effects of adding
breaks to a single sublattice of our system could not explain our observed results.
Here, we consider the opposite case in which equal length breaks are added to both sublattices
of the system. As such, the effective indices of both waveguides are the same, as is the added
loss. This amounts to a `trivial' addition of non-Hermiticity to the system, in the form of
$i \tau \hat{I}$ within Eq.\ (\ref{eq:s3}) rather than $i \tau \hat{\sigma}_1$, and thus should
\textit{not} result in the formation of a Weyl exceptional ring. To confirm this, we show simulations
of both the transverse radial propagation, $\langle \psi | r_\perp | \psi \rangle$, and the
output power ratio, $P_A / P_B$, in Fig.\ \ref{fig:ddrop}, for waveguide arrays with breaks added
to both sublattices. This approximately doubles the loss per unit cell, so we halve the lengths
of the breaks in these systems relative to those studied in the main text with breaks added to only
a single sublattice.

As can be seen in Fig.\ \ref{fig:ddrop}(a), this uniform addition of loss
to the system preserves the Weyl point of the system, as exhibited by a peak in the transverse radial
propagation. The slight shift in the Weyl point's wavelength is due to the slight shift in the
effective indices of the waveguides of the system, which changes the effective coupling constant
between the waveguides, but which remain in tune as breaks have been added
to both waveguides. Moreover, as loss has been added to both sublattices, the output power
remains equal in both sublattices for all of the simulated break lengths, as shown in Fig.\ \ref{fig:ddrop}(b).
These simulations provide further proof that adding breaks to only one sublattice of the helical
waveguide array results in the formation of a Weyl exceptional ring.

\end{document}